\documentclass[12pt]{iopart}
\usepackage{iopams}

\newcommand{\im}{\mathop{\mathrm{Im}}}
\newcommand{\re}{\mathop{\mathrm{Re}}}

\newcommand{\erfc}{\mathop{\mathrm{erfc}}}

\begin{document}

\title{Schur function averages for the real Ginibre ensemble}

\author{  Hans-J\"{u}rgen Sommers$^1$ and Boris A Khoruzhenko$^2$}

\address{$^1$ Fachbereich Physik, Universit\"{a}t
Duisburg-Essen, 47048 Duisburg, Germany }

\address{$^2$ Queen Mary University of London, School of Mathematical Sciences, London E1 4NS, UK }

\eads{\mailto{H.J.Sommers@uni-due.de},
\mailto{b.khoruzhenko@qmul.ac.uk}}

\begin{abstract}

We derive an explicit simple formula for expectations of all Schur functions in the real Ginibre ensemble.
It is a positive integer for all entries of the partition even and zero otherwise. The result can be used to
determine the average of any analytic series of elementary symmetric functions by Schur function expansion.

\end{abstract}


\section{Introduction }

Real asymmetric random matrices, and in particular statistics of their (complex) eigenvalues, have many interesting applications in modelling of a wide range of physical phenomena. They appeared in the studies of stability of complex biological networks \cite{May72}, dynamics of neural networks with asymmetric synaptic couplings \cite{Sompolinsky88}, directed quantum chaos in randomly pinned superconducting vortices \cite{Efetov97}, delayed time series in financial markets \cite{KWAPIEN06}, quantum chromodynamics with a real representation of the Dirac operator\cite{HOV}, or random quantum operations in quantum information theory \cite{Bruzda08}.

All statistical information about eigenvalues of random matrices is
encoded in their joint probability density function (pdf). For real
asymmetric random matrices the joint pdf of eigenvalues was derived
in \cite{Lehmann91} almost 20 years ago. It gave insights about the
generic properties of the eigenvalue distribution, like the level
repulsion, but had a complex structure due to singularities which
made the precise statistical inference about eigenvalues difficult
at that time. It suffices to say that the eigenvalue counting
measures for complex \cite{Edelman97} and real eigenvalues
\cite{EKS94} were initially obtained without the aid of knowing the
joint pdf. Very recently the interest in the spectral properties of
real asymmetric random matrices has been revived \cite{Kanzieper05}
and the Pfaffian representation of $n$-point eigenvalue correlation
functions has been derived directly from the joint pdf by various
methods \cite{Sinclair06,Forrester07,Sommers2007,Borodin08}. This
progress made it possible to compute the bulk and edge scaling
limits of the eigenvalue correlation functions \cite{Borodin08a}
and, in particular, to recover the results of \cite{Edelman97,EKS94}
about the eigenvalue counting measures in the limit of infinite
matrix dimension \cite{Forrester07,Sommers2007}. Interestingly, the
kernel that determines the eigenvalue correlation functions can be
written in terms of averages of the characteristic polynomial of
random matrices of lower dimension, much in the spirit of
\cite{Edelman97,EKS94}, see also \cite{FKh07} for complex matrices,
and hence can be computed in a very simple way \cite{APS}. One
intriguing open question is the relation of real random matrices to
integrability theory such as $\tau$-functions which are generating
functions for the cumulants of power sums of eigenvalues.

The eigenvalues of the real asymmetric matrices are either real or
pairwise complex conjugate. This implies a complicated and singular
structure of the eigenvalue correlation functions. It seems that in
many aspects the elementary symmetric functions of eigenvalues,
$e_n$, or their first N power sums, $t_n$, form a more convenient
set of independent real variables. The Jacobian of the
transformation from this set to the eigenvalues is just the
Vandermonde determinant appearing in the joint pdf of eigenvalues
\cite{Lehmann91}. Since symmetric functions of eigenvalues can be
expanded in products of $e_n$'s (or $t_n$'s), averages of symmetric
functions of eigenvalues can be written in terms of the joint
moments of elementary symmetric functions (or power sums).
Equivalent to all these moments is the set of all averages of Schur
functions which form another basis in the space of symmetric
functions. As Schur functions are determinants, divided by the
Vandermonde their average can again be reduced to a Pfaffian. This
Pffafian can be evaluated explicitly as we will show in this paper.
The result is very simple. If the partition characterizing the Schur
function is odd then the Schur function average is zero. For even
partitions the Schur function average is a positive integer
expressible in terms of Gamma functions. This gives an alternative
method of evaluating averages of symmetric functions of eigenvalues
by the way of Schur function expansions.

We will treat both even and odd matrix dimensions $N$ in a coherent
way. We will reveal a further interesting structure of the theory.
The basic building block is a skew symmetric kernel ${\cal
K}_N(z_1,z_2)$, which is a polynomial of degree $N-1$ in both
eigenvalues $z_1,z_2$, thus building a skew symmetric from with a
matrix $A^{-1}$. If one knows this matrix $A^{-1}$ in all even
dimensions, one is able to determine all correlation functions and
all moments. Since $A^{-1}$ has the same form in all dimensions,
which just can be considered as a submatrix of one infinite
dimensional matrix,  the   averages of general Schur functions are
given by Pfaffians of finite submatrices of this infinite
dimensional matrix.  For odd dimension one has to modify this
procedure slightly.

\section{The real Ginibre ensemble}
In this paper we consider real random $N\times N$ matrices $H$ with probability measure
\begin{equation}
\label{dmuH} d\mu (H) =  \exp \left( - \frac{1}{2} \sum_{i,j=1}^{N}
H_{ij}^{2} \right) \ \prod_{i,j=1}^{N} \frac{dH_{ij}}{\sqrt{2 \pi}}.
\end{equation}
This ensemble of random matrices is known under the name of real
Ginibre ensemble \cite{Ginibre65}. The matrix distribution
(\ref{dmuH}) induces a probability distribution on the eigenvalues
of $H$ which is known to be of the form
\cite{Lehmann91,Edelman97,SW}
\begin{equation}\label{dmu}
      d\mu(z_1,z_2, \ldots, z_N) = C_N
     \cdot \prod_{i<j} (z_i - z_j) \cdot \prod_k f(z_k)  \cdot  dz_1 \ldots dz_N,
\end{equation}
with positive function $ f(z)=f(\bar z)>0$, $ f^2(z)= e^{-\re
(z^2)}\erfc (|\im z|\sqrt{2}). $
The eigenvalues $z_j$ of $H$ are real or pairwise
complex conjugate and are ordered in such a way that $d\mu \ge 0$.

Introducing the skew symmetric weight function ($z=x+iy,$ $d^2z=dxdy$)
\begin{eqnarray}
\nonumber
 {\cal F} (z_1, z_2)= f(z_1)f(z_2) ( 2 i \delta^2 (z_1 - \bar{z}_2)  {\rm sgn} (y_1-y_2)  + \delta(y_1) \delta(y_2)  {\rm sgn} ( x_2-x_1))
\end{eqnarray}
with $ \delta^2(z_1-\bar{z_2})=\delta(x_1-x_2)\delta(y_1+y_2)$, all correlation functions are determined with the help of the skew symmetric kernel
\begin{equation}
\label{KN1}
 {\cal K}_N (z_1, z_2) = \sum_{k,l=1}^{  N}  {A}^{-1}_{kl}  z_1^{k-1} z_2^{l-1}
\end{equation}
with the skew symmetric matrix
\begin{equation}
\label{A} A_{kl} = \int d^2z_1 \int d^2z_2 \; {\cal F} (z_1,z_2)
z_1^{k-1} z_2^{l-1}\ .
\end{equation}
The eigenvalue correlations are Pfaffians with combinations of
${\cal K}_N$ and ${\cal F }$ as entries \cite{SW}. Comparing the
$1$-point density $R_1(z)=\int d^2 u {\cal F} (z, u) {\cal K}_N (u,
z)$ with Edelman's expression \cite{Edelman97} for the density of
complex eigenvalues,
\[
R_1(z) = f(z)f(\overline{z}) \frac{z-\overline{z}}{i\sqrt{2\pi}}
\sum_{n=0}^{N-2} \frac{(z\overline{z})^{n}}{n!}, \quad \im z >0 ,
\]
one finds
\begin{equation}
\label{KN}
{\cal K}_N (z_1,z_2) = \frac{z_1 - z_2}{2 \sqrt{2 \pi}} \sum_{n=0}^{N-2} \frac{(z_1 z_2)^n}{n!} \; .
\end{equation}
This result can independently be found by a simple supersymmetric
(Grassmannian) calculation of the average of a product of two
characteristic polynomials. This calculation can be extended to
partly symmetric or chiral counterparts \cite{APS}.

With the help of the duplication formula for the Gamma function
\begin{equation}
2\sqrt{2\pi} \Gamma(N-1)=2^{(N-1)/2}\Gamma((N-1)/2)\cdot 2^{N/2}\Gamma(N/2)
\end{equation}
(\ref{KN1}) and (\ref{KN}) imply that
\begin{equation}
\label{A-1}
A^{-1}_{kl} = a^{-1}_k \epsilon_{kl}a^{-1}_l
\end{equation}
with $a_k=2^{k/2}\Gamma (k/2)$ and $\epsilon_{kl}$ being the
tridiagonal skew symmetric matrix with $-1$ in the upper diagonal.
The structure of the Vandermonde determinant in (\ref{dmu}) and (or)
the results of paper \cite{SW} imply that this expression can be
used for odd dimension too, if we extend the matrix $A^{-1}$ to
dimension $N+1$ by putting $a_{N+1}=1$ in (\ref{A-1}). In this case
the skew symmetric matrix $\epsilon$ can be inverted and the inverse
yields (\ref{A}) from (\ref{A-1}). However, in that case formula
(\ref{KN1}) holds with the sum over $k,l$ running only from $1$ to
$N$. The normalization constant $C_N$ in ({\ref{dmu}) is given by
\begin{equation}
\label{1/C}
1/C_N = {\rm Pfaff}(A)
\end{equation}
that is the Pfaffian of the skew symmetric matrix $A$.

\section{Schur function averages}

The description of the system by the complex eigenvalues might not
be the most convenient one. A more suitable set of independent real
variables is the set of power sums
\begin{equation}
\label{tn}
t_n=\Tr(H)^n=\sum_{i=1}^N z_i^n
\end{equation}
or the set of elementary symmetric functions $e_1,\dots ,e_N$ which
are the coefficients in the expansion of the characteristic
polynomial in powers of $x$,
\begin{equation}
\label{en}
\det(1+xH)=\sum_{n=0}^N x^n e_n(z_1, \ldots, z_N)\ .
\end{equation}
Sometimes complete symmetric functions $h_n$ come in handy. These
are the coefficients in the expansion of the reciprocal
characteristic polynomial in powers of $x$,
\begin{equation}
\label{b2}
1/\det(1-xH)=\sum_{n=0}^{\infty}x^n h_n(z_1, \ldots, z_N)\ .
\end{equation}
Homogeneous symmetric polynomials in $z_i$ of degree $\ge 1$ can be
expressed as sums of products of power sums $t_n$, or equally as
sums of products of $e_n$'s or, equally, of  $h_n$'s. Hence, when
calculating averages of the symmetric polynomials in eigenvalues of
$H$, or analytic functions, one is led to the problem of finding the
joint moments of the eigenvalues. A full description of all
symmetric moments of eigenvalues can be given by the so-called Schur
polynomials or Schur functions. These functions are indexed by
sequences, called partitions, $\lambda=(\lambda_1, \ldots,
\lambda_N)$ of ordered non-negative integers $\lambda_1\ge
\lambda_2\ge\dots\ge\lambda_N\ge0$, called parts, and can be
conveniently written as the ratio of two determinants
\begin{equation}
\label{sigma} \sigma_{\lambda}(z_1, \ldots, z_N)=
\det(z_m^{N-n+\lambda_n})/\det (z_m^{N-n})\; .
\end{equation}
We will follow the convention of not showing zero parts, i.e.
\[
(\lambda_1, \lambda_2, \ldots, \lambda_l, 0,\ldots,0)\equiv (\lambda_1, \lambda_2, \ldots, \lambda_l)
\]
and when referring to specific partitions will adopt the notation
$\lambda=(k^{a_k}, \ldots , 2^{a_2}, 1^{a_1}) $ to indicate that
there are precisely $a_1$ of 1's, $a_2$ of 2's, etc., among the
parts of partition $\lambda$.

It can be verified directly from (\ref{sigma}) that if partition
$\lambda$ consists of one part, $\lambda=(n)$, then the
corresponding Schur function is just the complete symmetric
function, $\sigma_{(n)}=h_n$ and if partition $\lambda$ has only
zeros or ones among its parts,   $\lambda =(1^n)$ then the
corresponding Schur function is just the elementary symmetric
function, $\sigma_{(1^n)}=e_n$. Also, by convention,
$\sigma_{\lambda}=0$ is the number of non-zero parts of $\lambda$ is
greater than the number of indeterminants $z_j$.

If  $\lambda$ is a partition of $n$, i.e. $\lambda_1+\ldots
+\lambda_N=n$, then the Schur function $\sigma_{\lambda}$ is a
homogeneous symmetric polynomial of degree $n$. In fact the set of
all such Schur functions form a basis in the space of homogeneous
symmetric polynomials of degree $n$. In particular, for $n\ge 1$,
see, e.g., \cite{Sagan91},
\begin{equation}\label{b3}
t_n=\sum_{k=1}^n (-1)^{n-k} \sigma_{(k,1^{n-k})}\; ,
\end{equation}
where the sum is over all hook partitions $\lambda=(k,1^{n-k})$ of
$n$.  Vice versa, as already mentioned above, every Schur function,
as any other symmetric polynomial, can be expressed in terms of
products of power sums. One consequence of this is that the Schur
function of matrix argument defined by $\sigma_{\lambda}
(H)=\sigma_{\lambda} (z_1, \ldots, z_N)$  with $z_j$ being the
eigenvalues of $H$, is a polynomial in the matrix entries of $H$ and
$\sigma_{\lambda}(H)=\sigma_{\lambda}(THT^{-1})$ for any
non-degenerate matrix $T$.

The Schur functions are the characters of irreducible representations of the unitary group and as such are orthogonal with respect to integration over the Haar measure. This property is quite useful for finding the coefficients in Schur function expansions. For example, expanding the product of characteristic polynomials in Schur functions,
\[
\prod_{j=1}^{n} \prod_{k=1}^{N} (1+x_jz_k) = \sum_{\lambda}
c_{\lambda} (x_1, \ldots, x_n) \sigma_{\lambda} (z_1, \ldots, z_N),
\]
one can find the ``Fourier coefficients'' $ c_{\lambda} (x_1, \ldots, x_n)$ by integration over the unitary group
\begin{equation}\label{b4}
c_{\lambda} (x_1, \ldots, x_n) = \int \prod_{j=1}^{n} \det (I+x_jU)\
\overline{\sigma_{\lambda} (U)} dU,
\end{equation}
where the integration is over the unitary group $U(N)$  with respect
to the normalized Haar measure $dU$. The group integral in
(\ref{b4}) can be easily evaluated yielding $c_{\lambda}=\det
(e_{\lambda_{k}-k+j})$ which, by the Jacobi-Trudi formula is again a
Schur function,  $\det
(e_{\lambda_{k}-k+j})=\sigma_{\lambda^{\prime}}$, with
$\lambda^{\prime}$ being the partition conjugate to
$\lambda$\footnote{The conjugate to partition $\lambda=(\lambda_1,
\lambda_2, \ldots )$ is the partition
$\lambda^{\prime}=(\lambda^{\prime}_1, \lambda^{\prime}_2,\ldots)$
with $\lambda^{\prime}_j$ being the number of $\lambda_k$ that are
greater than or equal to $j$. }. Thus one arrives at the Schur
function expansion
\begin{equation}\label{b5}
\prod_{j=1}^{n} \prod_{k=1}^{N} \det (1+x_jH)= \sum_{\lambda}
\sigma_{\lambda^{\prime}}(x_1, \ldots, x_n) \sigma_{\lambda} (H).
\end{equation}
This expansion is well known under the name of dual  Cauchy identity
and can be derived by purely algebraic means, see e.g.
\cite{Macdonald95}. Other known Schur function expansions can be
obtained via (\ref{b4}), see, e.g., Appendix in \cite{FKh07a}.

The Schur function averages over the real Ginibre ensemble
(\ref{dmuH}) can  be obtained by the same reasoning as in
(\ref{A-1}) -- (\ref{1/C}) or following the calculations in
\cite{SW}. The only difference is that the Vandermonde
$\det(z_m^{N-n})$ has to be replaced by $\det(z_m^{N-n+\lambda_n})$.
The result is again a Pfaffian,
\begin{equation}
\label{s(z)}
\langle \sigma_{\lambda}(H)\rangle_N \propto {\rm Pfaff}\int d^2z_1d^2z_2 {\cal F} (z_1,z_2) z_1^{N-n+\lambda_n} z_2^{N-m+\lambda_m}
\end{equation}
where the brackets mean average over the real Ginibre ensemble (\ref{dmuH}) and $n,m=1,2\dots N$. The normalization constant can be restored requiring that $\langle \sigma_{\lambda}(H)\rangle_N=1$ for the empty partition $\lambda=(0)$. By making use of  (\ref{A}) and (\ref{A-1}), one can also write
\begin{equation}
\label{s(z)1}
\langle \sigma_{\lambda}(H)\rangle_N \propto      {\rm Pfaff} A _ {N-n+\lambda_n+1,  N-m+\lambda_m+1}\ .
\end{equation}
The indices $N-n+\lambda_n+1$ form an increasing sequence in the opposite direction:
\begin{equation}
\label{sequence}
1+\lambda_N <2+\lambda_{N-1}<\dots <N+\lambda_1 \ .
\end{equation}
Here it does not matter that $N-m+\lambda_m +1$ can be $>N$ because it turns out that $A_{l,m}$ has the same form in all dimensions. The reason is that $(\epsilon)^{-1}_{n,m}$ can be obtained from any $M$ dimensional matrix $\epsilon_{n,m}$ with even $M\ge n,m$. This is due to the special tridiagonal structure, as one can easily check (see also the following).  Therefore, and this is the simple idea, we can immediately calculate all
\begin{equation}
\label{Alm}
A_{l,m} = a_l\  (\epsilon^{-1})_{l,m}\ a_m
\end{equation}
which then is valid also for any submatrix centered along the diagonal. Again for odd dimension $N$ we obtain $\epsilon^{-1}$ from a higher even dimension say $M$ cutting all rows and columns with number not equal $N-n+\lambda_n +1$ or $M$ and put $a_M=1$. This is meant-and correspondingly in the even $N$ case- if we write $\epsilon^{-1} _ {N-n+\lambda_n+1,  N-m+\lambda_m}$.  Thus
\begin{equation}
\label{s(z)2}
\langle \sigma_{\lambda}(H)\rangle_N \propto      {\rm Pfaff} (a_{N-n+\lambda_n+1}\epsilon^{-1} _ {N-n+\lambda_n+1,  N-m+\lambda_m+1}a_{N-m+\lambda_m+1})\ .
\end{equation}
To calculate the Pfaffian, let us define it by a Grassmann integral (or Berezin integral)
\begin{equation}
\label{Pfaff}
 {\rm Pfaff}(A_{l,m})=\int d\chi_1\dots d\chi_N \exp(-{1\over 2}\sum_{l,m}\chi_l A_{l,m} \chi_m  )
\end{equation}
with anticommuting variables $\chi_i$. Then the Berezinian (Jacobian) for the transformation $\chi_l\to \chi_l/a_l$ is $\prod_l a_l=\prod_{l=1}^Na_l$ (note that it appears with the opposite exponent as compared to commuting variables). Thus we obtain
\begin{equation}
\label{s(z)3}
\langle \sigma_{\lambda}(H)\rangle_N \propto   (\prod_{n=1}^N a_{N-n+\lambda_n+1} )  {\rm Pfaff} ( \epsilon^{-1} _ {N-n+\lambda_n+1,  N-m+\lambda_m+1})\ .
\end{equation}
For calculating the matrix $(\epsilon^{-1}_{k,l})$ we may choose the smallest even dimension $M$, that contains all $N-n+\lambda_n+1$ : $M\ge N +\lambda_1+1  $.

Let us illustrate the matrix $\epsilon$, which we always use in even dimension
\begin{equation}
\label{epsilon}
\epsilon\ \ \ =\pmatrix{0&-1&0&0&0&0&\cdots\crcr
                  +1&0&-1&0&0&0&\cdots\crcr
                  0&+1&0&-1&0&0&\cdots\crcr
                  0&0&+1&0&-1&0&\cdots\crcr
                  0&0&0&+1&0&-1&\cdots\crcr
                  0&0&0&0&+1&0&\cdots\crcr
                  \vdots&\vdots&\vdots&\vdots&\vdots&\vdots&\ddots}
 \end{equation}
It is easy to see that  ${\rm Pfaff}(-\epsilon)=\int d\chi_1\dots d\chi_N(-\chi_1 \chi_2)(-\chi_3 \chi_4)\dots =1$
The corresponding inverse $\epsilon^{-1}$ has the form
\begin{equation}
\label{epsilon-1}
\epsilon^{-1}=\pmatrix{0&+1&0&+1&0&+1&\cdots\crcr
                  -1&0&0&0&0&0&\cdots\crcr
                  0&0&0&+1&0&+1&\cdots\crcr
                  -1&0&-1&0&0&0&\cdots\crcr
                  0&0&0&0&0&+1&\cdots\crcr
                  -1&0&-1&0&-1&0&\cdots\crcr
                  \vdots&\vdots&\vdots&\vdots&\vdots&\vdots&\ddots}
 \end{equation}
Again we see that  ${\rm Pfaff}(\epsilon^{-1})=\int d\chi_1\dots d\chi_N(-\chi_1 \chi_2)(-\chi_3 \chi_4)\dots =1$. Now let us calculate the Pfaffian in (\ref{s(z)3}). We first try to calculate the determinant which is simpler (remember $({\rm Pfaff} A )^2=\det A $), then we determine the sign. To calculate subdeterminants of $\epsilon^{-1}$ we consider the generating function
\begin{equation}
\label{DN}
D_N=\det(x_k\delta_{k,l} +(\epsilon^{-1})_{k,l})\ .
\end{equation}
From this we can calculate subdeterminants by differentiating with respect to $x_k$ to cut row and column number $k$ and then put $x=0$. Now we transform using $\det\epsilon^{-1}=\det\epsilon=1$
\begin{equation}
\label{DN1}
D_N=(\prod_{k=1}^Nx_k)\det(\frac{1} {x_k}\delta_{k,l} +\epsilon_{k,l})\ .
\end{equation}
which now we may generalize to odd $N$ and deduce due to the simple form of $\epsilon$
\begin{equation}
\label{DN2}
D_N= D_{N-1} + x_N x_{N-1} D_{N-2}
\end{equation}
with $D_1=1,\ \ D_2=1+x_1x_2$. As a result
\begin{equation}
\label{DN3}
D_N=1+x_1x_2+x_2x_3+\dots + x_1x_2x_3x_4+\dots
\end{equation}
is a sum of products of consecutive pairings with all coefficients equal to $+1$. This implies that we can cut only consecutive pairs. As a result: $ |{\rm Pfaff}(\epsilon^{-1}_{N-n+\lambda_n+1, N-m+\lambda_m+1})|=1$, if all $\lambda_n$
 even and $=0$ if one or more $\lambda_n$ odd. To determine the sign we observe that a permutation of rows and columns of the Pfaffian changes it by the sign of the permutation. Let us move in this way all rows and columns, which we don't want to cut, into the upper left corner of the matrix. This yields a sign $(-1)^{|\lambda|}$ of the permutation (which is $=+1$ if $|\lambda|=\lambda_1+\lambda_2+\dots+\lambda_N$ is even). Then we observe, that if we move away only consecutive pairs of rows and columns, the remaining submatrix of $\epsilon^{-1}$ in the upper left corner does not change its form. Thus its Pfaffian is $+1$. As the final result we obtain
\begin{equation}
\label{s(z)4} \langle \sigma_{\lambda}(H)\rangle_N=
\cases{\displaystyle{2^{\frac{|\lambda|}{2}}\prod_{n=1}^N
\frac{\Gamma((N-n+\lambda_n+1)/2)}{\Gamma((N-n +1)/2)}} & if all
$\lambda_n$ are even\\ 0 & otherwise.}
\end{equation}
Here $|\lambda|=\lambda_1 +\dots+\lambda_N$ and the upper limit $N$
of the product can also be  replaced  by the length of the
partition, which is the number of its nonzero parts $\lambda_n$. It
has to be mentioned that the Schur function average (\ref{s(z)4}), can
be also obtained from zonal polynomials \cite{Macdonald95}, however
this derivation is rather indirect.


\section{Conclusions }

We have calculated the average of Schur functions over the real Ginibre ensemble (\ref{dmuH}) by using the Pfaffian representation (\ref{s(z)2}) for the average and then evaluating the Pfaffian explicitly. It vanishes for odd partitions and is given by a simple expression in terms of Gamma functions (\ref{s(z)4}) for even partitions. This result might be useful for evaluating averages of symmetric polynomials and analytic functions in eigenvalues of real Gaussian matrices. For example, consider the power sums $t_n$, $n\ge 1$. Their averages can easily be calculated by the way of Schur function expansion (\ref{b3}) and our main result (\ref{s(z)4}). For odd $n$ there are no hook partitions of $n$ with all parts even and therefore the average of $t_n$ vanishes. Of course, this can be verified directly from the symmetries in the matrix distribution (\ref{dmuH}). If $n$ is even then there is only one hook partition of $n$ with all parts even. This partition is $\lambda=(n)$. Hence, all terms but one on the right-hand side in (\ref{b3}) vanish and one arrives at the relation
\[
\langle \Tr H^n \rangle_N = \langle \sigma_{(n)}(H)\rangle_N =\langle h_{n}(H)\rangle_N
\]
where $h_n$ is the complete symmetric function of degree $n$. Hence, by (\ref{s(z)4}),
\[
\langle \Tr H^{2m} \rangle_N = 2^{m} \frac{\Gamma(N/2+m)}{\Gamma(N/2)}=\prod_{j=1}^{m}(N+2(m-j)), \quad m\ge 1.
\]
Similarly, on averaging (\ref{b5}) with the help of (\ref{s(z)4}) one obtains
\[
 \langle \prod_{j=1}^{n} \det (1+x_jH)\rangle_N= \sum_{n/2\ge \lambda_1\ge \lambda_2 \ge \ldots \ge \lambda_N \ge 0} \!\!\! c_{\lambda}\  \sigma_{(2\lambda)^{\prime}}(x_1, \ldots, x_n),
\]
where $\lambda=(\lambda_1, \ldots, \lambda_N)$, $2\lambda=(2\lambda_1, \ldots, 2\lambda_N)$ and
$
c_{\lambda}= 2^{|\lambda|}\prod_{k}\frac{\Gamma(\lambda_k+ (N-k+1)/2)}{\Gamma((N-k+1)/2)}.
$ If $n=2$ then the constraints $n/2\ge \lambda_1\ge \lambda_2 \ge \ldots \ge \lambda_N \ge 0$ imply that $\lambda_j=1$ or $0$ for each $j$. Correspondingly, $\lambda=(1^k)$, $2\lambda =(2^k)$ and $(2\lambda)^{\prime} =(k,k)$, where $k$ is the number of non-zero $\lambda_j$. Observing that $\sigma_{(k,k)}(x_1,x_2)=(x_1x_2)^k$ and $c_{(1^k)}=N!/(N-k)!$ one concludes that
\begin{equation}\label{b6}
\langle \det (1+x_1H)\det (1+x_2H) \rangle_N = N! \sum_{k=0}^N \frac{(x_1x_2)^k}{(N-k)!},
\end{equation}
in agreement with calculations in \cite{Edelman97} and \cite{APS}. This expansion can also be obtained via (\ref{en}) directly from the symmetries in the ensemble distribution (\ref{dmuH}). One only needs to recall the fact that the elementary symmetric polynomial of degree $n$ in the eigenvalues of $H$ is the sum of all principal minors of $H$ of order $n$. Then (\ref{b6}) follows from the invariance of the distribution (\ref{dmuH}) with respect to changing the sign of $H_{lm}$ and, also, permutation of columns of $H$.

\ack
The first author acknowledges support by SFB/TR12 of the Deutsche Forschungsgemeinschaft.


\section*{References}
\begin{thebibliography}{10}
\bibitem{May72} R.M.May, Nature {\bf 298 } 413 (1972)
\bibitem{Sompolinsky88} H.Sompolinsky, A.Crisanti and H.-J.Sommers, Phys. Rev. Lett. {\bf 61} 259 (1988)
\bibitem{Efetov97} K.B.Efetov, Phys. Rev. Lett. {\bf 79} 491 (1997)
\bibitem{KWAPIEN06} J.Kwapien, S. Drozdz, A.Z. Gorski and  F. Oswiecimka, Acta  Phys. Pol. {\bf B37}, 3039 (2006)
\bibitem{HOV} M.A.Halasz, J.C. Osborn, and J.J.M. Verbaarschot,Phys. Rev. {\bf D56}, 7059(1997)
\bibitem{Bruzda08} W.Bruzda, V.Cappelini, H.-J.Sommers, K.\.Zyczkowski,  Phys Lett {\bf A 373} 320 (2009)
\bibitem{Lehmann91} N.Lehmann and H.-J. Sommers, Phys. Rev. Lett. {\bf 67}, 941 (1991)
\bibitem{Edelman97} A.Edelman, J. Multivariate Anal. {\bf 60}, 203 (1997)
\bibitem{EKS94} A.Edelman, E.Kostlan and M.Shub, J. Amer. Math. Soc. {\bf 7}, 247 (1994)
\bibitem{Kanzieper05} E.Kanzieper and G. Akemann, Phys. Rev. Lett. {\bf 95}, 230201 (2005)
\bibitem{Sinclair06} C.D.Sinclair, Int. Math. Res. Not., {\bf 2007}, 1 (2007).
\bibitem{Forrester07} P.J.Forrester and T. Nagao, Phys. Rev. Lett. {\bf 99}  050603 (2007)
\bibitem{Sommers2007} H.-J. Sommers, J. Phys. A {\bf 40}, 671  (2007)
\bibitem{Borodin08} A.Borodin, C.D.Sinclair, arXiv: math-ph/0706.2670v2
\bibitem{Borodin08a} A.Borodin, C.D.Sinclair, arXiv: math-ph/0805.2986
\bibitem{FKh07} Y. V. Fyodorov and B.A. Khoruzhenko, Comm. Math. Phys.  {\bf 273}, 561 (2007).
\bibitem{APS} G.Akemann, M.J. Phillips, and H.-J. Sommers, J.Phys.A {\bf 42} 012001 (2009)
\bibitem{Ginibre65} J.Ginibre, J. Math. Phys. {\bf 6}, 440 (1965)
\bibitem{SW} H.-J. Sommers and W. Wieczorek, J.Phys. {\bf A41}, 405003 (2008)
\bibitem{Sagan91} B. E. Sagan, The Symmetric Group, 2nd ed. (Springer) 1991
\bibitem{Macdonald95} I.G. Macdonald, Symmetric Functions and Hall Polynomials, 2nd ed. (Clarendon Press) 1995
\bibitem{FKh07a} Y. V. Fyodorov and B.A. Khoruzhenko, J. Phys. A {\bf40} 669 (2007)
\bibitem{Mehta04} M.L.Mehta, Random Matrices (Amsterdam Elsevier) 2004
\end{thebibliography}
\end{document}